# NUTRIENT REMOVAL FROM AGRICULTURAL RUN-OFF IN DEMONSTRATIVE FULL SCALE TUBULAR PHOTOBIOREACTORS FOR MICROALGAE GROWTH

Joan García[1*], Antonio Ortiz[1], Eduardo Álvarez[1], Vojtech Belohlav[1,2], María Jesús García-Galán[1], Rubén Díez-Montero[1], Juan Antonio Álvarez[3] and Enrica Uggetti[1]

[1]GEMMA-Environmental Engineering and Microbiology Group, Department of Civil and Environmental Engineering, Universitat Politècnica de Catalunya-BarcelonaTech, c/ Jordi Girona 1-3, Building D1, E-08034 Barcelona, Spain

[2]Department of Process Engineering, Czech Technical University in Prague, Technicka 4, 166 07 Prague, Czech Republic

[3]AIMEN Technology Centre, c/ Relva 27 A, Torneiros, E-36410 Porriño, Pontevedra, Spain

*Corresponding author:

Tel.: +34 934016464

Fax. +34 934017357

E-mail address: joan.garcia@upc.edu

**Abstract**

The objective of this paper is to present the design, construction and operation (during one year) of 3 full scale semi-closed, horizontal tubular photobioreactors (PBR, 11.7 m$^3$ of volume each) used to remove nutrients of a mixture of agricultural run-off (90%) and treated domestic wastewater (10%). PBRs were located outdoor and have 2 paddlewheels (engines of 0.25 kW) to ensure the movement of the mixed liquor. The microalgal biomass produced in the PBRs was harvested in a static lamella settling tank in which a polyaluminium chloride coagulant is applied. Each PBR treated in average 2.3 m$^3$/d, being the actual mean hydraulic retention time 5 d. PBRs were submitted to strong seasonal changes regarding solar radiation and temperature, which had a direct impact in the activity of microalgae and the efficiency of the system. Higher mixed liquor pH values were registered in summer (daily average > 10). These high values were not observed in the effluents because the system was designed to discharge the mixed liquor (effluent) only at the end of night, when pH reached the lowest daily values (around 8.5). Most of the influent and effluent nitrogen content was inorganic (average of 9.0 mg N/L and 3.17 mg N/L, respectively), and in the form of nitrate (62% and 50%, respectively). Average nitrogen removal efficiency was 65%, with values of around 90% in summer, 80% in autumn, 50 % in winter and 60% in spring. Most of the influent and effluent phosphorus content was in the form of ortophosphate. Influent average was 0.62 mg P/L, but with great variations and in a considerable number of samples not detected. Removal efficiency (when influent values were detected) was very high during all the study, usually greater than 95%, and there were not clear seasonal trends for efficiency as observed for TIN. Volumetric biomass production greatly changed between seasons with much lower values in winter (7 g VSS (volatile suspended solids)/m$^3$·d) than in summer (43 g VSS/m$^3$·d). Biomass separation efficiency of the settler was very good in either terms of turbidity and total suspended solids, being most of the time lower than 5 UNT and 15 mg/L, respectively. Overall this study demonstrated the reliable and good effectiveness of microalgae based technologies such as the PBR to remove nutrients at a full scale size.

1. **Introduction**

Changes in the nutrient biochemical flows due to anthropogenic activities are one of the main environmental challenges that humanity must face in the coming decades. The alteration of the cycles of nitrogen and phosphorus (N and P) is already considered of high global risk, with unfavourable effects leading to unknown impacts (Steffen et al., 2015). Urban and agricultural discharges of contaminated or insufficiently treated water are the main cause for the imbalance of these biochemical cycles. Nowadays, most of the aquatic ecosystems are receiving these nutrient enriched discharges, being their eutrophication an unequivocal signal of it. Globally, more than 450 coastal areas are affected by severe eutrophication (Selman et al., 2008).

Ecological engineering techniques can be used to reverse this contamination situation in many cases, allowing also for the restoration of these aquatic ecosystems. In particular, treatment wetlands have been extensively used in recent decades as effective systems for the treatment of urban, agricultural and even industrial wastewater; a vast array of literature with hundreds of examples at full scale is available (Ávila et al., 2013; García et al., 2010). There is much less experience with other types of ecological engineering techniques such as microalgae systems, despite the fact that microalgae based wastewater treatment systems were developed more than 50 years ago, specifically to treat urban wastewaters (García et al., 2006). Therefore, it is necessary to show and demonstrate the potential of these microalgae technologies at full-scale. One of the most powerful advantage of microalgae systems in comparison to other technologies is that harvested microalgae biomass can easily be valorised as a bioproduct and/or energy, which is extremely interesting within the framework of the circular economy.

Up to date, most of the studies devoted to phytoremediation of agricultural-related wastes by means of microalgae have focused on lab-scale experiments to treat industrial effluents, such as those from dairy farms (Labbé et al., 2017), palm oil mills (Kamyab et al., 2015) or rice mills (Kumar et al., 2016). The treatment of aquaculture effluents and diluted pig slurry treatments were also investigated in different works (Ansari et al., 2017; Lananan et al., 2014; Ledda et al., 2016). The capacity of microalgae to remove pesticides from agriculture run-off was also evaluated by Matamoros et al. (2016). In all cases, however, only lab-scale experiments were performed, usually with microalgae cultures grown on synthetic media and aseptic conditions. To the authors' knowledge, only two recent studies have evaluated the feasibility of integrating agricultural run-off treatment and biomass production at real scale. Bohutskyi et al., (2016) studied the phytoremediation of agricultural run-off by filamentous green microalgae (*Cladophor*a sp. and *Rhizoclonium* sp.) in an Algal Turf Scrubber (ATS®), treating 10 million gallons per day. The authors obtained a maximum monthly productivity of 22 g/m$^2$·d (measured as volatile suspended solids) and a suitable feedstock to obtain biogas after anaerobic digestion. Furthermore, diluted digestate from anaerobic digestion was used as nutrients supplement to cultivate more valuable microalgae species. The second study by García-Galán et al. (2018) evaluated the efficiency of a large-scale photobioreactor treating agriculture run-off and also obtaining microalgae biomass as added-value product. Results showed a maximum biomass production of 76.4 g/m$^3$·d (measured as total suspended solids) in April, and a total N elimination ranging from 84% to 95%.

In the present paper we describe the experience gained on the design, construction and operation during the first year after the start-up (from May 2017 to May 2018) of 3 full scale photobioreactors (PBRs) fed with a mixture of agricultural run-off and treated domestic wastewater. The microalgae biomass produced in the photobioreactors was harvested in a static lamella settling tank. All these units were constructed in the framework of the innovation European project INCOVER (http://incover-project.eu/). These PBRs are part of a complex installation aiming to efficiently treat wastewater and produce bioenergy, bioproducts and reclaimed water for irrigation. A brief description of the entire experimental site can be found in Uggetti et al., (2018). This study is exclusively focused on the PBRs functioning and their auxiliary elements. The INCOVER project will be operative till May 2019.

2. **Materials and Methods**

2.1. **Photobioreactors design**

The PBRs are located in the Agròpolis experimental campus of the Universitat Politècnica de Catalunya-BarcelonaTech (UPC) (41.288 N, and 2.043 E UTM), very near to Barcelona's airport (Figure 1 and Figure A1 in the Appendixes). The PBRs and their auxiliary elements were conceived, designed and constructed by the GEMMA Research Group of the UPC in collaboration with the company Disoltech S.L after several previous investigations (García-Galán et al., 2018; Solimeno et al., 2017; Uggetti et al., 2018). These PBRs are tubular horizontal semi-closed reactors, each consisting of 2 lateral open tanks made from 10 mm polypropylene (5 m width, 1 m length and 0.6 m height, nominal volume of 1.25 $m^3$ each at design water depth). Both tanks are connected through 16 low density polyethylene tubes (0.3 mm thick, 125 mm diameter and 47 m length, nominal volume of 9.2 $m^3$ for all tubes together) (Figure 2 and Figure S2). These tubes lie down on a plastic covering sheet in order to ensure separation from the ground, and they are protected by agricultural anti-birds nets. The total useful volume of each PBR is 11.7 $m^3$ (approximately 20% corresponding to the tanks, and 80% to the tubes). In each open tank, a paddlewheel with eight blades (1 m width, 0.35 m long) is installed 1.8 m away from the external edge and at 3 cm height from the bottom. An engine (0.25 kW) connected to each paddlewheel provides a turning speed which can be changed from 0 to 12 rpm. Rotation of the paddlewheel makes the mixed liquor contained in the tank move from a shallow water sector to a deep one. Difference in pressure head causes a gravity flow through 8 tubes from the deep side of one tank to the shallow side of the opposite one. Then again, the flow is moved by the paddlewheels to the deeper side part of the tank, and then it returns to the shallow side of the first tank through the other 8 tubes, and so on (Figure S2). Each tank has an inclined dam in the deep sector, which assists in maintaining the two different surface water levels and avoids big waves within the tank (Figure 2). Both open tanks ensure and favour the homogenous distribution and mixing of the liquor and also the release of the exceeding dissolved oxygen accumulated along the closed tubes.

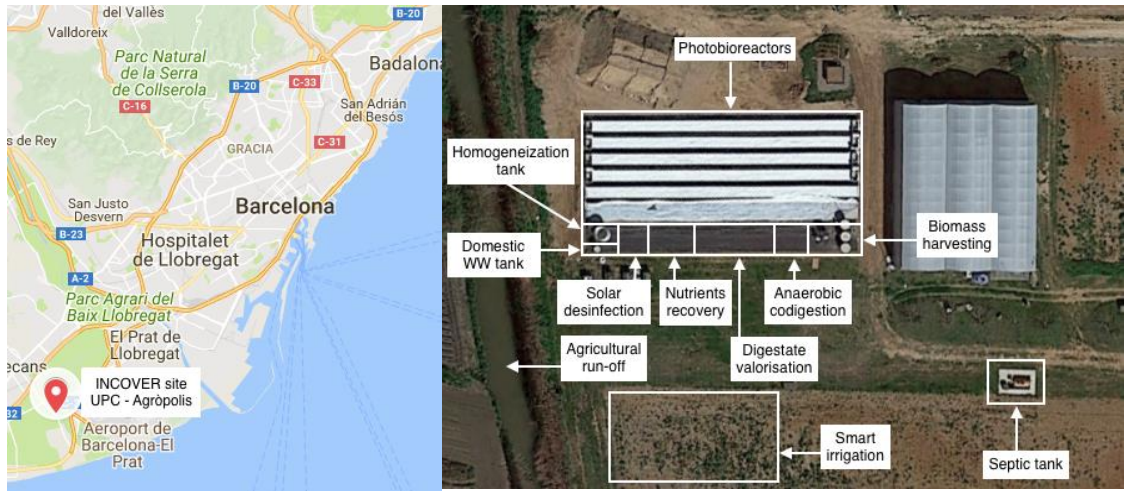

**Figure 1.** Location map of the facilities (left) and plan view of unit processes in INCOVER project site (right).

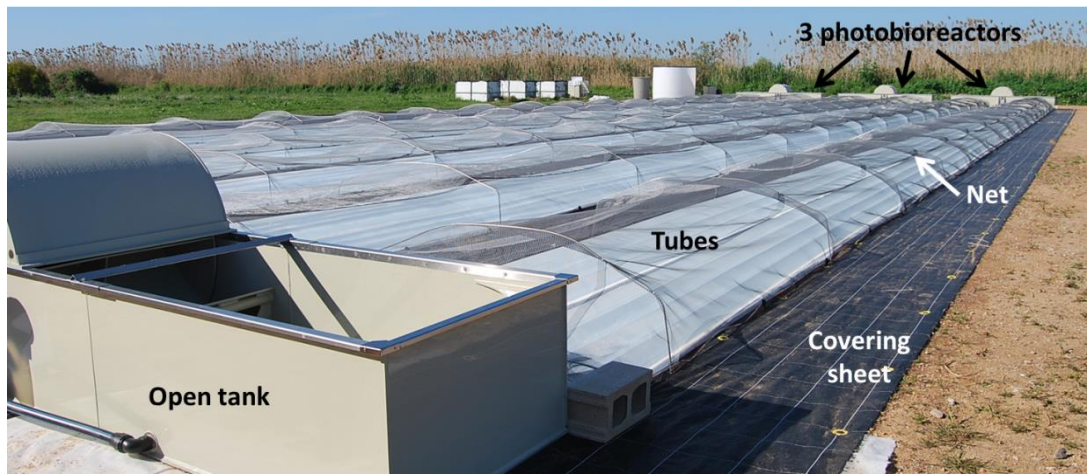

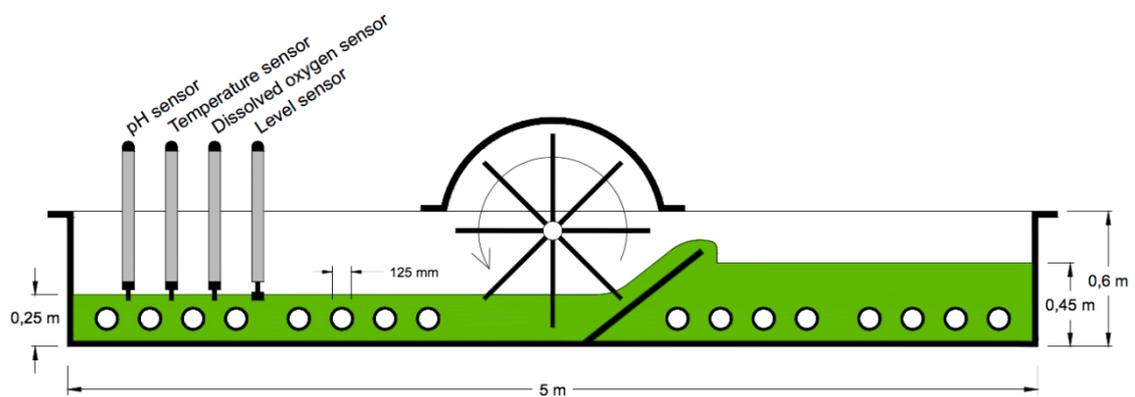

**Figure 2.** Picture of the 3 photobioreactors located in parallel showing the tubes and the open tanks (upper) and lateral schematic representation of one tank (lower). Note the paddlewheel with the eight blades (which has a cover), and the inclined dam.

Each PBR is equipped with online sensors of pH (Hatch Lange SL., Spain), dissolved oxygen (DO) (Neurtek, Spain) and temperature (Campbell Scientific Inc., USA) in one of the two open tanks. Data of these parameters are taken every 5 s and recorded and stored each 60 s in a datalogger (Campbell Scientific Inc., USA). PBRs also include a water level sensor (Wras, UK) to control filling and emptying operations. They also have an automatic $CO_2$ injection system (tubing, valves and pressure sensor), but at the time of the present work it was not being used.

The three PBRs were installed in winter 2016-17, and were inoculated at the end of April 2017 with a mixed culture grown in experimental high rate algal ponds fed with urban wastewater (Gutiérrez et al., 2016). A volume of 10 L was added to each PBR, with a volatile suspended solids (VSS) concentration of approximately 220 mg/L. The inoculum consisted of a community of bacteria, microalgae, protozoa and small metazoa, but mostly dominated by green microalgae *Chlorella* sp. and *Stigeoclonium* sp., and diatoms *Nitzschia* sp. and *Navicula* sp. (Gutiérrez et al., 2016). Note that *Stigeoclonium* is a branched filamentous microalgae which in natural aquatic environments usually grows attached to submerged surfaces. In the particular case of the inoculum used here, it was growing in the form of flocs submerged in the mixed liquor of the high rate algal ponds. After inoculation, the three PBRs were operating in parallel and fed with a mixture of agricultural run-off and domestic wastewater (design ratio of 6:1 respectively, although the actual ratio was slightly higher). Total design flow was 7 $m^3$/d (6 $m^3$/d of agricultural run-off and 1 $m^3$/d of treated domestic wastewater) (see Supplementary Materials, Methods section).

### 2.2. Auxiliary elements description and system operation

Treated domestic effluent is obtained from an aerated septic tank which receives the wastewater of the main building of the campus Agròpolis (~20 persons, without overnight stay), whereas agricultural wastewater comes from a drainage collection channel (see Figure 1). Figure 3 shows a process flow diagram of the PBR and their auxiliary elements. Daily operation cycle starts at 4:30 AM when the treated domestic wastewater, stored in a cylindrical glass fiber tank (TK-103, 1 $m^3$) discharges in a cylindrical polyethylene homogenization tank (HT-102, 10 $m^3$, provided with a sampling port) through stream line 3. This operation is done by means of a centrifugal pump P-104 (14.4 m³/h) during a maximum time of 30 min. Treated domestic wastewater continuously reaches TK-103 by the stream line 2, that conveys treated wastewater to the tank thanks to a submersible pump located in the aerated septic tank. When more than 1 m³ domestic wastewater is produced per day, TK-103 remains full and the remaining wastewater gets out through a weir, reaching a general by-pass. Note that the amount of treated domestic wastewater is often lower than 1 m³ (i.e., during the weekends).

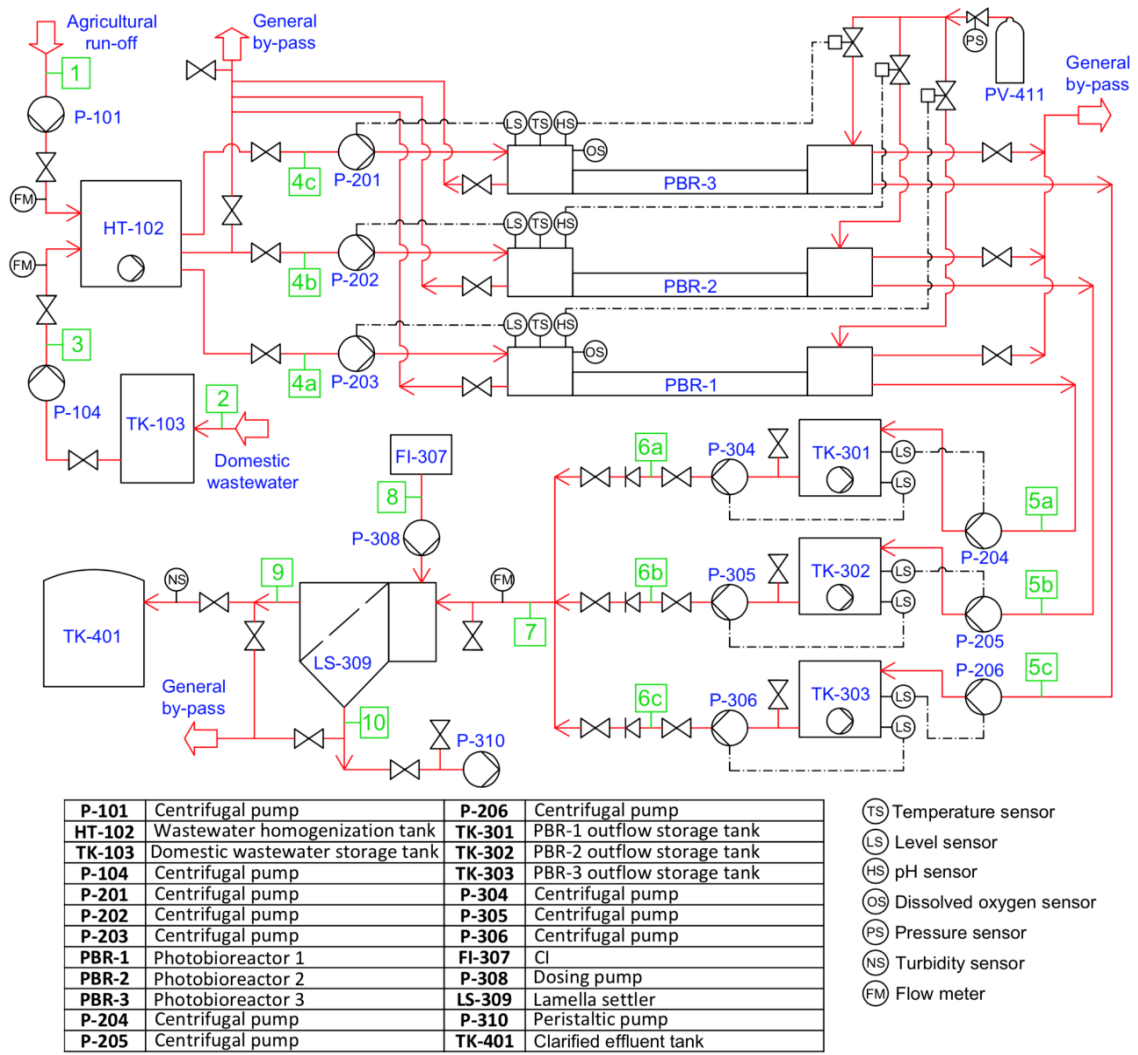

**Figure 3.** Process flow diagram of the photobioreactors and their auxiliary elements.

At 5:00 AM, 6.0 m³ of agricultural run-off are pumped by a submersible pump (P-101), from the nearby channel to the homogenization tank during a period of approximately 3 hours (stream 1). Agricultural run-off and domestic wastewater are mixed in that tank and discharged in a unique outflow at 8:00 AM (streams 4a, 4b and 4c) to the 3 PBRs by means of 3 centrifugal pumps (P-201, P-202 and P-203, time and level (LS) controlled) during a period ranging from 1 to 1.5 h. The homogenization tank has an internal recirculation pump to ensure the complete stirring of the tank. Note that stream lines 1 and 3 have flowmeters (FM) installed (Siemens, Germany). The homogenization tank has the necessary tubing and valves for maintenance and discharge to the general by-pass.

At 7:00 AM, and before filling the PBRs with fresh wastewater, a designed volume of 2.3 $m^3$ of mixed liquor of each PBR are simultaneously discharged to 3 circular glass fiber outflow storage tanks (TK-301, TK-302 and TK-303, 2.7 $m^3$ each) through stream lines 5a, 5b and 5c. This operation is done by 3 centrifugal pumps (P-204, P-205 and P206) which provide a constant flow of approximately 3 $m^3$/h. Therefore, that volume is evacuated from the PBRs daily in 45 min approximately. Each outflow storage tank has two different water level sensors (LS, located at the top and at the bottom) that control the filling and emptying pumps, as well as internal submerged pumps for complete stirring of the tanks and different sampling ports.

As a result of the operation procedure described above, PBR functioning is done in parallel and in semi continuous mode. Note that the emptying and filling processes are done early in the morning (once a day) in order to promote biomass growth when sunlight is available. PBR theoretical hydraulic retention time (HRT) is 5.0 d. PBRs are usually operated with a paddlewheel speed ranging from 10 to 12 rpm, providing a depth in the shallow water sector of 0.25 m, and 0.45 m into the deep sector (Figure 2). This 0.2 m difference in pressure head gives a theoretical hydraulic speed of 0.25 m/s inside the tubes, which ensures turbulent flow (Reynolds number, Re >30,000 at 20 °C). HRT within the tubes at design speed is 3.25 min, and 0.81 min in the two tanks (both together). A single drop of water theoretically gives daily 360 loops inside the PBR, and the paddlewheels move more than 4,200 $m^3$ of water every day. PBRs have the necessary tubing and valves installed for maintenance and discharge to the general by-pass.

Water from the storage tanks reach the polypropylene static lamella settler (LS-309, 0.675 m$^3$, Moral Torralbo S.L.U., Spain) through stream lines 6a, 6b, 6c and 7, the last one provided with a sampling port (Figure 4). The apparent critical settling velocity within the settler is 1.14 m/h at a flow of 0.4 m$^3$/h. Peristaltic pumps P-304, P-305 and P-306 (0.4 m$^3$/h) work one after the other, starting at 8:00 AM, 3:00 PM and 10:00 PM (storage tanks are therefore emptied in this order). A flowmeter (FM, Siemens, Germany) in stream line 7 allows checking the flowrate to the settler. The settler has one internal mixing chamber (0.05 m$^3$) for coagulation-flocculation where polyaluminium chloride (PAX-18 (Al$^{3+}$ at 9%), Kemira Water Solutions Inc., Spain) is added at a concentration that usually ranges from 2 to 10 mg Al$^{3+}$/L, by means of a diaphragm pump (stream line 8, P-308), which is activated simultaneously with the peristaltic pumps. Internal stirring of this chamber is provided by a motorized propeller (0.09 kW). The coagulant is stored in a plastic tank (FI-307, 0.1 m$^3$). The water is discharged from the lamella chamber of the settler by stream line 9, provided with an online turbidimeter (Digimed, Brazil), to one clarified effluent tank (TK-401), and from there to other units which are not described for the purposes of this study (Figure A1). The coagulant doses are adjusted by means of jar-tests using the methods described in Gutiérrez et al. (2015), which are carried out when it is observed that the effluent turbidity from the settling tank is higher than 5 UNT. The biomass harvested in the settler is taken out through stream line 10 with a peristaltic pump (P-310) to the other units neither described here (Figure S1). There is a sampling port for the harvested biomass, and the settler has the necessary tubing and valves for maintenance and discharge to the general by-pass.

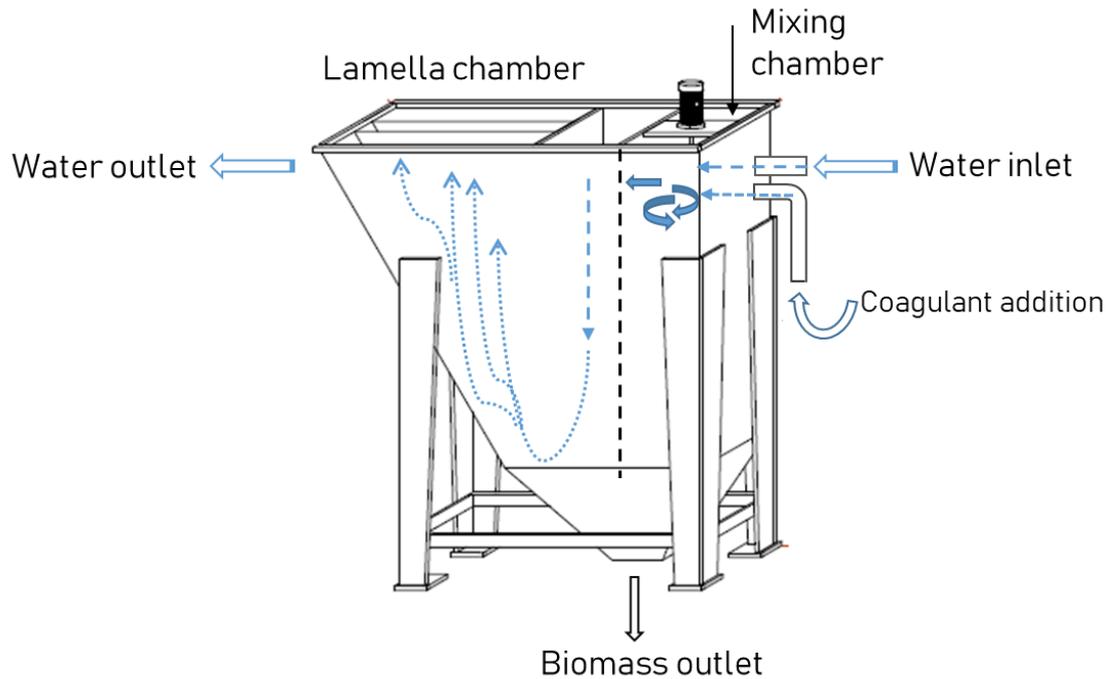

**Figure 4.** Schematic representation of the static lamella settler used for biomass harvesting.

### 2.3. Samples and analyses

Weekly grab samples were taken from May 2017 to May 2018 (every Tuesday at 8:00 AM) from the homogenization tank (influent), and each of the 3 outflow storage tanks (effluent, which is actually the mixed liquor of the PRBs), transported to the laboratory and immediately analysed for temperature, pH, electrical conductivity, turbidity, alkalinity, nitrate, nitrite, ammonia, orthophosphate, chemical oxygen demand (COD), total suspended solids (TSS) and volatile suspended solids (VSS). Effluent COD was analysed from filtered samples to avoid influence of microalgae biomass. Samples were not taken during August in midsummer. Nitrate, nitrite and orthophosphate were analysed using an ion chromatograph DIONEX ICS 1000 (Thermo-scientific, USA). Ammonia was analysed using the methods described in Solorzano (1969). Alkalinity, COD, TSS and VSS were analysed with using the procedures indicated in the book *Standard Methods for the Examination of Water and Wastewater* (APHA-AWWA-WPCF, 2001). Temperature and electrical conductivity were measured with an Endress+Hauser equipment (Germany), while turbidity and pH were measured with Hanna instruments (USA). COD, TSS and VSS were always analysed in triplicate. N removal was studied considering total inorganic nitrogen (TIN, sum of nitrate, nitrite and ammonia), because previous studies in the same site demonstrated that organic nitrogen was too low to have a significant importance in mass-balance calculations (Uggetti et al., 2016). Note that in these studies organic nitrogen was usually lower than 1 mg N/L, representing less than 10% of the total nitrogen. P removal was studied using orthophosphate because other species of phosphorus were insignificant. Note that in this investigation, due to the size of the facilities, the main objective was to evaluate the efficiency of the PBRs for nutrient removal during wastewater treatment, rather than the evaluation in depth of nutrient transformations and pathways. Also note that the settling tank was put in operation and optimized from the end of November 2017, and therefore data are available only from December 2017. TSS and turbidity of the settling tank were analysed from two to three times per week. Meteorological information was obtained from Catalonian Meterogical Service: www.meteo.cat/observacions/xema/dades.

Mixed liquor samples were occasionally observed at the microscope Nikon Eclipse E200 (Japan), provided with an epi-fluorescence illuminator. This microscope was equipped with a camera (Fi2, Nikon, Japan) connected to a computer (software NIS-Element viewer®).

3. **Results and discussion**

Changes in total daily flow as recorded in stream lines 1 and 3 (Figure 3) are shown in Figure 5. Table 1 shows the average total daily flow, and the average daily flow of agricultural run-off and domestic wastewater. The total flow was quite similar to the design flow (7.0 m$^3$/d, considering the three PBRs), and the relative proportion of agricultural run-off and treated wastewater wass also similar to the design ratio (6:1). The observed agricultural run-off flow approximately represents in average a 90% of the total flow. Most of the variations of the total flow were due to fluctuations in domestic wastewater flow, which during weekends was very low or even 0. Punctual extreme values were usually related due to small disarrangements in the corresponding pump. According to the average daily flow (2.3 m$^3$/d per each PBR), the actual average HRT within the PBRs hydraulic retention time was 5 d, which is the same than the design value.

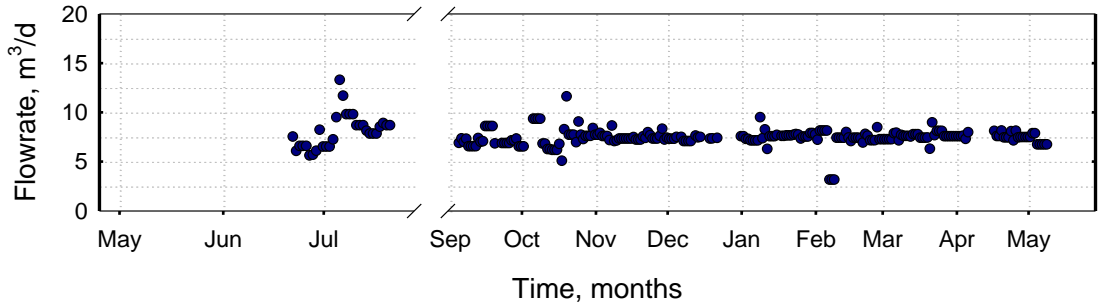

**Figure 5.** Changes in total daily flow to the three photobioreactors. Measurements started in July when flowmeters were installed. In August photobioreactors were stopped.

Table 1. Average daily flow of agricultural run-off and domestic wastewater, and average total flow. SD and ranges also shown.

| Type of wastewater | Average | SD | Range |
|---|---|---|---|
| Domestic wastewater, $m^3/d$ | 0.6 | 0.5 | 0-4.5 |
| Agricultural run-off, $m^3/d$ | 7.0 | 1.1 | 2.3-12.3 |
| Total flow, $m^3/d$ | 7.6 | 1.2 | 3.1-13.2 |

The PBRs were submitted to seasonal changes regarding solar radiation and daily average air temperature, which affected daily average mixed liquor temperature (Figure 6). As it can be observed, the daily average temperature of the mixed liquor was usually lower than 15 °C from the end of November to February. Changes in environmental factors had a direct impact in the activity of microalgae, which in turn affected to general water quality parameters such as pH. Higher daily average pH values were registered in summer, part of autumn and spring, and lower in winter. In fact, pH values >10 are usually observed in sunniest and warmest months in microalgae based treatment systems (García et al., 2006). These high values were not observed in the effluents of the PBR because they were pumped out at the end of night, when pH had the lowest values (around 8.5, see below in next paragraph). Note that in January there was an increase in pH which was linked to a rise in air and mixed liquor temperature.

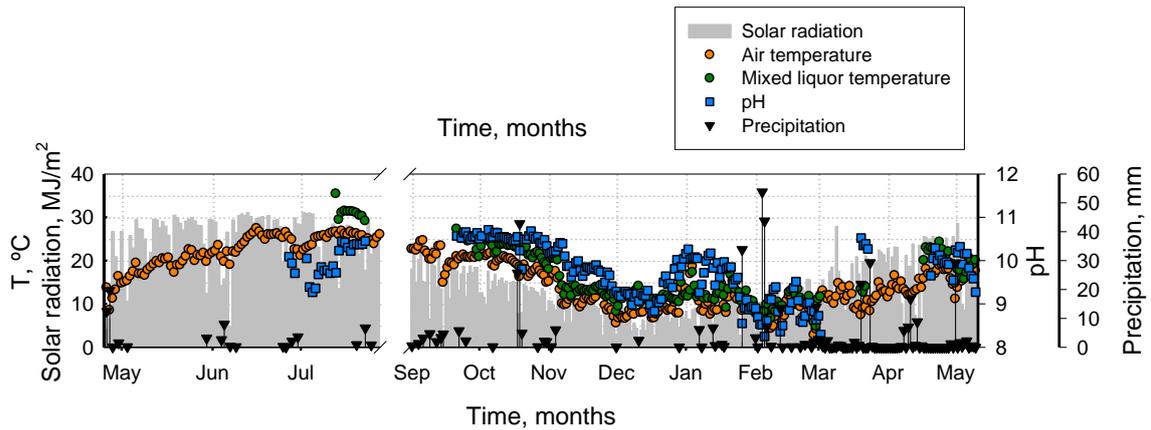

**Figure 6.** Changes in total solar radiation (absolute values), average air and photobioreactor mixed liquor temperature and pH (daily average values). Precipitation is also shown (absolute values). Mixed liquor temperature and pH were measured with on-line sensors installed in July. In mid-summer photobioreactors were stopped. Data of the mixed liquor corresponds to one of the three photobioreactors to avoid overlapping.

Diurnal changes in solar radiation and air temperature had a clear effect on mixed liquor temperature, pH, and to a lower extend in DO. Solar radiation and temperature had effects on algal activity (in particular on photosynthesis) which gave place to strong daily fluctuations of pH (Figure S3). These diurnal fluctuations are usually observed in microalgae based treatment systems such as high rate algal ponds. In fact, in these systems, DO fluctuations are stronger than those observed in this work because the constant organic load applied which gives place to DO values near 0 during night (García et al., 2006).

### 3.1. Nutrient removal

Table 2 shows the average values of the parameters measured weekly in the influent (homogeneization tank) and the effluent (in fact the mixed liquor) of the three PBRs (outflow storage tanks, TK-301, TK-302 and TK-303). Average effluent pH values were in the low range of the values shown in Figure 6, because the effluent was pumped out at the end of the night (8:00 AM) and stored in the outflow storage tanks, which were covered to avoid photosynthesis (values in Figure 6 are daily averages measured with sensors). Figure S3 shows how the lowest pH values are registered at the end of the night, due to the lack of photosynthesis in the dark and to the cellular respiration which is continuously taking place. Nevertheless, the pH was higher in the effluent than in the influent in all the samples. Alkalinity dramatically decreased from the influent to the mixed liquor in the PBRs due to the photosynthetic activity of the microalgae (inorganic C assimilation, which in turn affected $CO_3^-$ precipitation), which also led to the pH rise during the day.

**Table 2.** Average (±SD) and range of water quality parameters analyzed in influent and effluent photobioreactor samples taken weekly. Ranges also shown in brackets. Note that effluent averages are calculated from data of the three photobioreactors. Also note that the effluent in fact is the mixed liquor of the photobioreactors. "nd" means not detected. $n$= 45.

| Parameter | Influent | Effluent |
|---|---|---|
| **Temperature, °C** | 19.7 ±4.4 (12.8-26.9) | 19.6 ±4.8 (12.6-27.4) |
| **pH** | 8.21 ±0.24 (7.84-8.78) | 9.17 ±0.79 (7.57-10.79) |
| **Electrical conductivity, dS/m** | 2.67 ±0.43 (1.90-3.50) | 2.40 ±0.38 (1.31-3.10) |
| **Turbidity, NTU** | 40.4 ±18.5 (12.8-95.9) | 71.7 ±105 (6-445) |
| **Alkalinity, mg $CaCO_3$/L** | 280 ±34 (230-355) | 152 ±54 (94-392) |
| **Nitrite, mg N/L** | 0.86 ±1.46 (nd-6.55) | 0.42 ±1.05 (nd-8.21) |
| **Nitrate, mg N/L** | 5.53 ±3.34 (nd-11.65) | 1.57 ±2.50 (nd-12.88) |
| **Ammonia, mg N/L** | 2.53 ±2.88 (nd-13.26) | 1.21 ±3.75 (nd-25.46) |
| **Total inorganic nitrogen, mg N/L** | 9.00 ±5.01 (0.05-21.64) | 3.17 ±4.66 (nd-26.29) |
| **Ortophosphate, mg P/L** | 0.62 ±0.78 (nd-2.34) | 0.03 ±0.18 (nd-1.72) |
| **COD, mg/L** | 76 ±40 (12-160) | 91 ±56 (14-300) |
| **TSS, mg/L** | 66 ±78 (15-480) | 191 ±237 (4-1290) |
| **VSS, mg/L** | 21 ±16 (4-94) | 124 ±129 (2-550) |
| **VSS/TSS** | 0.38 ±0.13 (0.2-0.7) | 0.77 ±0.17 (0.2-1.3) |

Most of the influent TIN was generally in the form of nitrate (62%), while nitrite and ammonia remained in lower concentration (10% and 28%, respectively). Observed TIN concentration was clearly lower than expected during the design phase (see Methods section in the Supplementary materials). Effluent TIN was also dominated by nitrate (50% nitrate, 13% nitrite and 37% ammonia). Note that in the effluents of microalgae based treatment systems ammonia is usually low due to microalgae uptake as well as stripping linked to high pH values attained during the day (García et al., 2000).

TIN was therefore successfully removed with an annually global average of 65%, but removal efficiencies changed during the different seasons due to variations in environmental conditions. Efficiencies decreased from summer (91%) to autumn (83%), winter (49%) and spring (59%). Lower solar radiation and temperature in winter resulted into lower microalgae activity and growth, and higher concentrations of TIN in PBR mixed liquor (Figure 7). Also an atypical cold and rainy weather in a great part of spring gave place to not low concentrations in TIN PBR mixed liquor. García-Galán et al., (2018) evaluated the efficiency of a similar PBR treating the same water in spring (mostly in April) and obtained a TIN removal ranging from 84 to 95%.

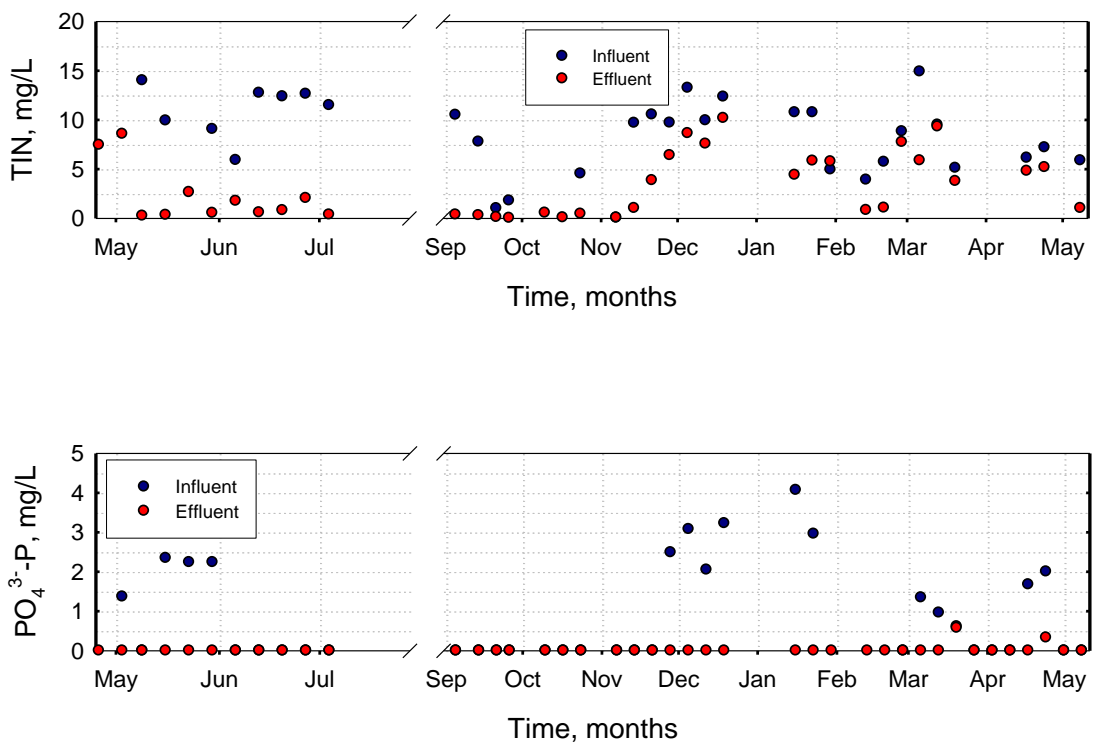

**Figure 7.** Changes in total inorganic nitrogen (TIN) and orthophosphate concentration in the influent and the effluent of the photobioreactors. Effluent values are the averages of the three photobioreactors. In mid-summer photobioreactors were stopped.

Similar to TIN, the concentration of orthophosphate in the influent was lower than that considered during the design phase of the system (see Methods section in Supplementary saterials). In fact, it was not detected in a considerable number of sampling events and the calculated removal efficiency (when influent values were detected) was very high during all the study, greater than 95%, showing not clear seasonal trends for efficiency as observed for TIN. This high removal was achieved due to the uptake by microalgae, but also to their indirect effect on rising pH and causing ortophosphate precipitation (García et al., 2002). Reasons behind the lower influent TIN and orthophosphate concentrations than those considered for design are not evident, since data used during the design phase were based on field measurements.

Average influent N:P ratio was 14.5:1 (in a milligram basis), slightly higher than that considered in design phase (7:1, see Methods section in the Supplementary Materials). This high ratio and the low phosphorus concentrations led mostly to a constant development of populations of Cyanobacteria in the PBRs, which were clearly dominant during summer and autumn. Most of the Cyanobacteria belonged to acoccal species resembling *Synechococcus* (Figure S4). Note that during these seasons, the influent N:P ratio was even higher than 20:1. In winter and a part of spring, however, Cyanobacteria were outcompeted by green microalgae. Microalgae of the inoculum were not observed in the PBRs after few months of operation.

### 3.2. Biomass production and final effluent quality

Microalgae biomass concentration and production was measured through solids analyses. Note that microalgae evaluation through pigment measurements (chlorophyll *a*) probably would give a more accurate measurement of biomass concentrations. However, in this work solids were measured for the sake of simplicity and because in these microalgae systems there is a very good correlation between chlorophyll *a* and solids (García et al., 1998, 2006). Also because microalgae production in culture systems is usually given in biomass weight (Ansari et al., 2017).

Solids concentrations were higher in the effluent (mixed liquor without biomass separation) than in the influent due to microalgae growth. In the effluent, average VSS represented the 78% of the TSS considering aggregated data of the three PBR. This percentage fits well into the values generally observed for microalgae based treatment systems (Gutiérrez et al., 2016), which are lower than other biomasses due to the high pH achieved in the mixed liquor, which promotes precipitation of inorganic salts of different nature. Biomass concentration showed great variations in the three PBRs, as evidenced by the high VSS standard deviation. For this reason, changes in the biomass concentration have been represented in box-plot charts for the different seasons (Figure 8). As observed, the biomass concentration is well related with environmental factors such as solar radiation and temperature. Considering the average concentration, the biomass production in the three PBR together was approximately 670 g VSS/d, which in volumetric terms represents approximately 20 g VSS/m$^3$·d (=20 mg VSS/L·d). Obviously, production greatly changed between seasons with much lower values in winter (240 g VSS/d, 7 g VSS/m$^3$·d) than in summer (1500 g VSS/d, 43 g/m$^3$·d). García-Galán et al. (2018) obtained very similar biomass production rates in a similar PBR treating the same wastewater. These authors registered the highest production in April (approximately 75 g TSS/m$^3$·d). Considering the VSS/TTS ratio found in the present work, the estimated production achieved by García-Galán et al. (2018) would be 57 g VSS/m$^3$·d, which is in the same orders of magnitude to the values observed in this study. The great variation of biomass concentration (and production) observed in spring (Figure 9) was linked to the atypical cold and rainy weather in part of February, entire March and part of April 2018.

Regarding the values observed for TIN and orthophosphate (Figure 6), it is quite clear that the biomass production was limited by nutrients during the studied period, except for winter and part of spring (when TIN concentration increased). Indeed, the environmental factors in corresponding months (solar radiation and temperature) are those that limit microalgae growth.

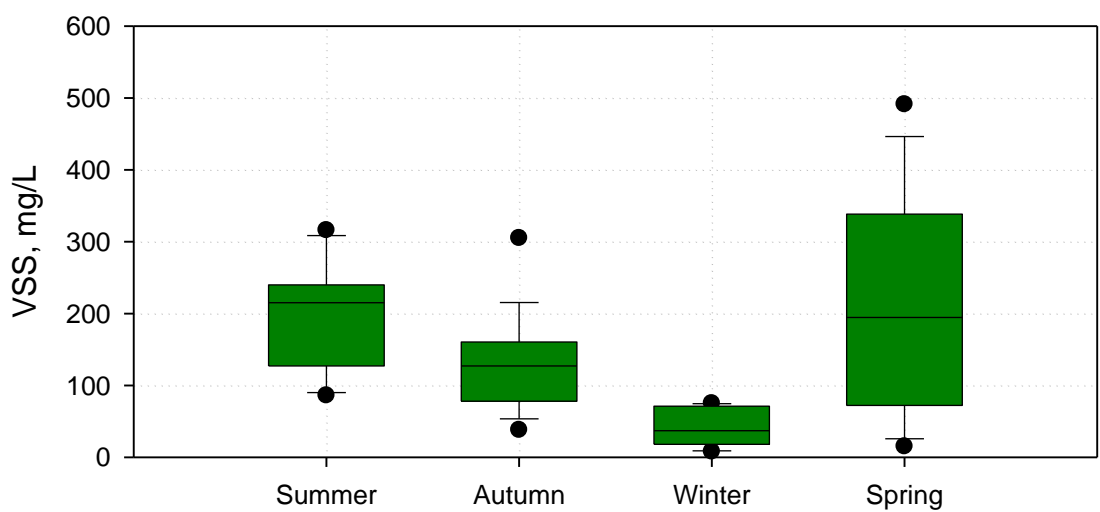

**Figure 8.** Box-plot of the average volatile suspended solids (VSS) concentration in the three photobioreactors in different seasons. The lower boundary of the box indicates the 25th percentile, the line within the boxmark the median (solid line), and the upper boundary of the box indicates the 75th percentile. Whiskers (error bars) above and below the box indicate the 90th and 10th percentiles, respectively. Upper and bottom dots represent the 95th and 5$^{th}$ percentile, respectively.

Influent COD values were relatively low due to the feedstock origin (mixture of agricultural run-off and treated domestic wastewater). PBR dissolved COD was slightly higher than total influent COD. This is due to the fact that, in these type of systems, microalgae biomass releases dissolved organic carbon (García et al., 2006). However, the impact of this increase in COD was irrelevant, as COD was clearly lower than the legal discharge requirements usually employed for secondary effluents (< 125 mg/L), together with the fact that TSS values in the final effluent were very low.

The final effluent was produced after microalgae biomass separation in the static lamella settler, which main task was water clarification. One of the main aims of the INCOVER project was to produce clarified water with a very low turbidity (< 5 UNT), in order to prevent the good performance of the units after the PBRs (not described in this work, but shown in Figure 1 and S1). Figure 9 shows the changes in turbidity values and TSS concentration. As it can be observed, the quality of the final effluent was very good, values being most of the time below 5 UNT, and only with few peaks exceeding it. These peaks were related to small failures in plant operation, such as coagulant depletion or inappropriate coagulant doses that had to be re-establish after jar-test evaluation. Note that coagulant dosed usually ranged from 2 to 10 mg $Al^{3+}$/L. The good quality in terms of turbidity was also reflected in the TSS concentrations, which were in all cases well below 25 mg/L, and in more than half of the cases below 15 mg/L. Note that dissolved constituents were scarcely changed from the PBR to the effluent of the settling tank.

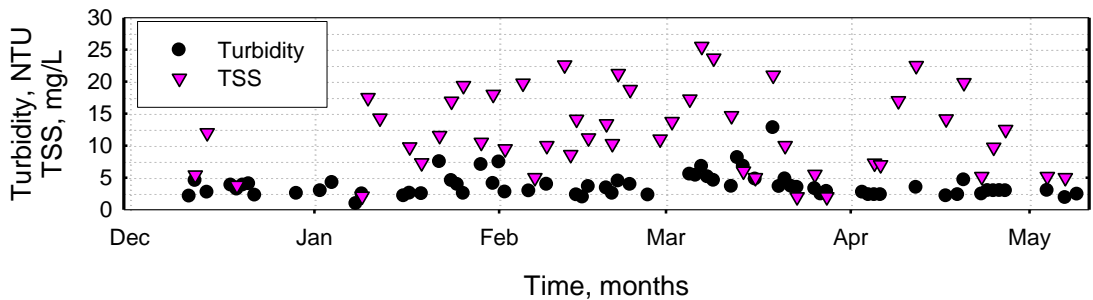

**Figure 9.** Changes in turbidity and TSS concentration after clarification (final effluent). Measurements from December 2018 when the settling tank was put in operation and optimized.

## 4. Conclusions

The present study demonstrates the efficiency of microalgae based technologies such as PBRs to remove nutrients from wastewater at full scale. Data of one year of operation of 3 full-scale semi-closed horizontal tubular PBRs (11.7 $m^3$) have been summarized and evaluated. These PBRs were conceived within the INCOVER project, with the dual purpose of removing nutrients from wastewater and producing biomass, which would be further valorized within a biorefinery concept and the circular economy framework. A daily flow of 2.3 $m^3$/d per PBR was treated, corresponding to a HRT of 5 d. Microalgae activity has a strong dependency on seasonal changes in solar radiation and temperature, which led to higher pH values in the mixed liquor detected in summer and great variations in biomass production between seasons, going from 7 g VSS/$m^3$·d in winter to 43 g VSS/$m^3$·d in summer. Biomass separation efficiency of the settler was very high, with turbidity and TSS values < 5 UNT and around 15 mg/L, respectively. Average nitrogen removal efficiency ranged from 50% in winter to 90% in summer. Phosphorus concentration was much lower than that of nitrogen in both influent and the clarified effluent, mostly corresponding to orthophosphate, and the removal efficiency was greater than 95% in almost all cases.

**Acknowledgements**

The authors would like to thank Mr. Javier Carretero and the environmental analysis laboratory of the GEMMA-UPC group for their uninterested help and support during this study. Also the authors would like to thank the European Commission [INCOVER, GA 689242] for their financial support. M.J. García and E. Uggetti would like to thank the Spanish Ministry of Industry and Economy for their research grants [FJCI-2014-22767 and IJCI-2014-21594, respectively].